\pdfoutput=1
\documentclass[5p, authoryear]{elsarticle}

\usepackage{lineno,subfig}
\usepackage{hyperref}
\usepackage{amsmath}

\journal{Industrial Journal of Industrial Ergonomics}

\biboptions{authoryear}

\begin{document}

\begin{frontmatter}

\title{A Study on Shared Steering Control in Driving Experience Perspective: How Strong and How Soon Should Intervention Be?}

% \author{Kyudong Park}
% \address{Radarweg 29, Amsterdam}
% \fntext[myfootnote]{Since 1880.}

\author[cite]{Kyudong Park}
\ead{kdpark@postech.ac.kr}

\author[ime]{Sung H. Han\corref{mycorrespondingauthor}}
\cortext[mycorrespondingauthor]{Corresponding author}
\ead{shan@postech.edu}

\author[cse]{Hojin Lee}
\ead{hojini33@postech.ac.kr}

\address[cite]{Dept. of Creative IT Engineering, POSTECH, Pohang, South Korea}
\address[ime]{Dept. of Industrial and Management Engineering, POSTECH, Pohang, South Korea}
\address[cse]{Dept. of Computer Science and Engineering, POSTECH, Pohang, South Korea}

\begin{abstract}
The lane keeping assistance system (LKAS), a representative of the advanced driver assistance system (ADAS), comprises a shared control that cooperates with the driver to achieve a common goal. The experience of the driver through the steering wheel may vary significantly depending on the steering control strategy of the system. In this study, we examine how driving experience changes according to various steering control strategies. Based on the preliminary study and typical LKAS parameters, nine control strategies (3 torque amounts (TOR) $\times$ 3 deviations to start control (DEV)) were designed as a prototype. Eighteen participants participated in evaluating each strategy in a highway environment provided by a driving simulator. Two-way repeated measure ANOVA was used to assess the effects of the system. Both the objective measures (standard deviation of lane position, steering reversal rate, and root mean square of lateral speed) and subjective measures (pleasure and arousal of emotion, trust, disturbance, and satisfaction) are analyzed. The experimental results demonstrate that all dependent measures are significant. As the TOR increased, SDLP decreased. However, no difference is observed between the 2-Nm and 3-Nm TOR in terms of trust and satisfaction. The high disturbance and negative emotion in 3 Nm appear to be the cause. In terms of the DEV, the high level of the root mean square of the lateral speed is observed at 0.8 m. Further, negative effects are found in pleasure, trust, and satisfaction. There is little difference at all dependent measures between 0.0-m and 0.4-m DEV. In the regression model analyzed from the aspect of satisfaction, the 2.32-Nm TOR and 0.27-m DEV are the optimal values. We expect our research on shared steering control with an assistance system to be applied to the experience design of a lateral semi-autonomous vehicle.

\end{abstract}


\end{frontmatter}

\section{Introduction}

Research to improve driver safety and convenience has been actively conducted worldwide based on the digitalization of automotive parts, miniaturization of sensors, and the development of computer vision technology. Under this trend, the advanced driver assistance system (ADAS), developed to help drivers recognize careless situations, are being actively introduced into vehicles \citep{bengler2014three}. A lane keeping assistance system (LKAS), a representative of the ADAS, is a system that aims to ensure the safety of the driver by preventing a vehicle from departing the lane through active intervention. The LKAS monitors the relative position between the vehicle and the lane using the front camera and sensors. In an unintentional lane departure, the LKAS maintains the lane by controlling the steering angle to reduce the risk of lateral collision accidents \citep{risack2000video, rajamani2012lateral, marino2012integrated}. It is the same as the lane departure warning system (LDWS) in that it is an assistance function for the lateral direction; however, the most significant difference is that the LKAS intervenes directly in the steering wheel operation. Thus, it shows the characteristics of the shared control \citep{mulder2015introduction}, in which the torque required by the system to avoid the departing lane and the driver’s steering torque are simultaneously generated. The intervention of the LKAS has altered the interaction between the vehicle \citep{eckoldt2012experiential} and the driver and has changed the driver’s decision-making model of steering wheel manipulation \citep{rodel2014towards}. Thus, it can be assumed that the driving experience will also be influenced by the interaction between the LKAS and the driver, considering that the driving task can be changed in the future owing to the introduction of the shared control system \citep{strand2014semi, cho2017technology}. Even if the assistance system installed in the driver’s vehicle secures safety with an unfavorable driving experience, the system may not be trusted and could be shut down. Therefore, it is important to design the LKAS control strategy considering the driver’s experience; nevertheless, empirical investigations into the driving experience of the LKAS control strategy have been scarce. \cite{eichelberger2016toyota} collected the actual driver experience of adaptive cruise control systems, forward vehicle warning systems, lane departure warning systems (LDWSs), and LKAS. Among them, they revealed that the LKAS caused the most annoyance among drivers. However, it relies on the driver’s memory and thus should not cause the annoyance thereof. \cite{kidd2017driver} suggested that the LKAS could be an alternative to the LDWS because it is not as annoying as the LDWS. Among the surveyed ADASs, however, the trust in LKAS was the lowest. \cite{reagan2017driver} analyzed the driver’s experience on actual production vehicles and observed that not all drivers deem that the LKAS improves their driving experience. They also observed that the driver’s sentiment varied by vehicle model. When we assume that different LKAS control strategies are applied to each vehicle model, it is important to study how the driving experience changes according to the control strategy. Preliminary studies \citep{park2018measuring, park2018modelling} revealed that an intrusive feeling could affect the driver’s satisfaction according to various LKAS parameters. It focused on the type of intrusive feeling felt by a few experts. Therefore, the correlation between the parameter setting of the LKAS and the driving experience still need to be identified through experiments with ordinary drivers.

This study aims to obtain the optimal LKAS control strategy with good driving experience. In Section 2, we first introduce the process of designing LKAS control strategies with two of the most influential parameters. The third section discusses the experimental method used for this study. The results of the experiments are summarized in the fourth section. We investigated how the driver’s experience changes with the strategies in terms of emotion, trust, disturbance, and satisfaction. Further, the driver’s behaviors under the distracted situation were collected and analyzed. The fifth section presents the findings of the research, focusing on the primary effects and regression analysis. Finally, the sixth section presents the conclusions and future works.

% 2장
\section{Designing LKAS control strategy}

The control of the LKAS is abstractly open to the public, and the detailed settings for the implementation are not disclosed, which is presumed to be a trade secret. Therefore, we developed a simplified LKAS prototype and applied it to our experiment. The process for designing LKAS control strategies comprises two steps. First, we empirically selected important factors influencing the driving experience among the LKAS parameters through the preliminary studies \citep{park2018measuring, park2018modelling}. Subsequently, several control strategies are designed by combining the selected factors in the LKAS prototype. 

\subsection{Determination of the influencing factors}

According to the preliminary study to obtain intrusive feelings in the testing vehicle with the LKAS, the largest number of negative phenomena (vibration of the steering wheel, heavy steering wheel, abrupt lateral change, and lane departure) were observed depending on the intervention torque amount (TOR) and deviation to start control (DEV). These two factors are considered in the control strategy.

\begin{itemize}
\item High TOR - Heavy steering wheel and abrupt lateral change
\item Low TOR - Lane departure
\item High DEV - Abrupt lateral change and lane departure
\item Low DEV - Vibration of the steering wheel
\end{itemize}

\begin{figure}
\begin{center}
  \includegraphics[width=0.7\linewidth]{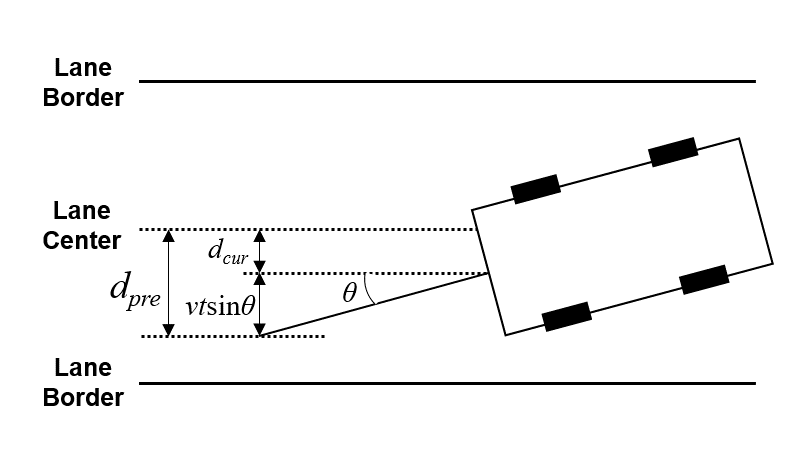}
  \end{center}
  \caption{Predicted deviation based on the lateral position and heading angle.}
  \label{fig:1}
\end{figure}

\begin{figure}
  \begin{center}
  \includegraphics[width=0.6\linewidth]{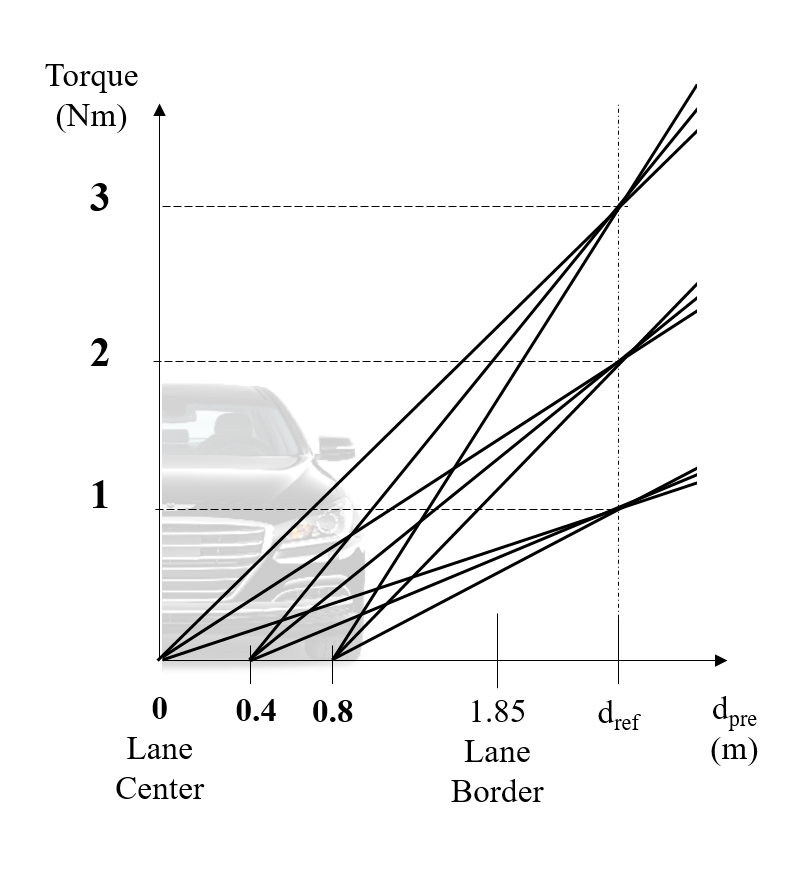}
  \end{center}
  \caption{Nine treatment conditions (steering control strategies) used in the experiment.}
  \label{fig:2}
\end{figure}

\subsection{Building of steering control strategy}

The LKAS control strategy should be designed from the operating principle of the LKAS. The simplified LKAS prototype model used in this experiment is designed to generate the amount of torque calculated by the predicted deviation ($d_{pre}$) after $t$ seconds, which is similar to the time-to-line crossing method \citep{pape1996dynamic, van2000comparison, mammar2006time}. However, it contains a different criterion for calculating the torque amount, i.e., it is based on the future position rather than on the distance from the lane border. The strict guidance, in which the torque is applied linearly according to the distance \citep[p. 694]{gayko2012lane}, is employed for a simple implementation and the easy interpretation of parameters. According to Fig. \ref{fig:1}, the predicted deviation $d_{pre}$ is calculated as follows:
\begin{equation}
    \label{eq:1}
    d_{pre} = vtsin\theta + d_{cur}
\end{equation}
where $v$ is the velocity of the vehicle (m/s); $\theta$ is the heading angle ($^\circ$); $d$ is the deviation (m) based on the current vehicle lateral location.

To generate the intervention torque amount (TOR) using $d_{pre}$, we defined the reference deviation $d_{ref}$ using eq. (\ref{eq:1}) and the following parameters: 1) heading angle of 1.72$^\circ$ derived from the lateral velocity of 0.6 m/s, which is higher than the NHTSA (National Highway Traffic Safety Administration) judgment condition of the LKAS (NHTSA, 2013), 2) deviation $d_{cur}$ of half the lane width, which is when the vehicle leaves the lane, and 3) $t$ of 1 s. When a vehicle’s velocity $v$ is 20 m/s and lane width $l$ is 3.7 m, which is the experimental condition in this study, $d_{ref}$ is approximately 2.45 m. The torque $T_{lka}$ generated by the LKAS is calculated as follows:
\begin{equation}
    \label{eq:2}
    T_{lka} = \frac{k_{TOR}}{(d_{ref}-k_{DEV} )} (d_{pre}-k_{DEV})
\end{equation}
where $k_{TOR}$ and $k_{DEV}$ are independent variables in this study. $k_{TOR}$ represents the amount of torque when $d_{pre}$ becomes the reference deviation $d_{ref}$, and $k_{DEV}$ is the deviation distance to commence the intervention. The larger the $k_{DEV}$, the later the torque intervenes when the drivers leave the lane center. Meanwhile, when $k_{DEV}$ becomes zero, the control torque always intervenes unless the driver maintains the position at the center of the lane perfectly. In Fig.~\ref{fig:2}, nine control strategies established by combining two parameters (3 TOR by 3 DEV) are shown. When $x$ becomes the $d_{ref}$ in each control line, the corresponding $y$ is the TOR. As shown, the $x$-intercept value in each control line is the DEV. 

\subsection{Hypotheses}

Considering the negative experiences caused by the TOR and DEV mentioned in Section 2.1, trust and disturbance are expected to be critical in the driving experience of the LKAS. For example, the abrupt lateral change and the lane departure will adversely affect one’s trust in the LKAS; further, the vibration of the steering wheel and the heavy steering wheel will affect the sense of disturbance. Depending on the TOR and DEV, a tradeoff may occur between trust and disturbance. If designers increase the TOR to secure the trust, the vehicle does not depart from the lane, but the driver may feel a significant amount of disturbance. By reducing the TOR to decrease the feeling of disturbance, the performance of the lane keeping is low and the trust might decline. The primary goal of this study is to identify the tradeoff as empirical data and to obtain the design parameters that exhibit the best driving experience considering the conflicts. The research hypotheses to be confirmed through the experiments are as follows:
\begin{itemize}
\item H1. The performance of the LKAS (SDLP, SRR, and RMSLS) significantly differs according to the LKAS control strategy (TOR $\times$ DEV).
\item H2. The driving experience (emotion, trust, disturbance, and satisfaction) significantly varies according to the LKAS control strategy (TOR $\times$ DEV).
\end{itemize}

\section{Experimental Methods}

\subsection{Participants}

Eighteen participants (11 males and 7 females) voluntarily participated in the experiment. Their ages ranged from 23 to 29 years (mean = 26.1, SD = 1.6). The average driving experience of the participants was approximately 37.8 months (SD = 26.6). Novice drivers who have less than 12 months’ driving experience were excluded from this experiment because the participants must have enough driving experience to evaluate the LKAS control strategies.

\subsection{Experimental design}

This experiment used a two-factor (i.e., TOR, DEV) within-subjects factorial design. The independent variables in this experiment were the torque amount (i.e., TOR, three levels), and deviation to start control (i.e., DEV, three levels). 

Several studies \citep{daun, vandanelzen2011} argued that the driver was overwhelmed and found it difficult to override the torque easily if the torque of the LKAS exceeded 3 Nm, Therefore, the level of the TOR in this experiment is set to 1 Nm, 2 Nm, and 3 Nm, which are within 3 Nm. The DEV level is set to 0.8 m, a distance in which the driver felt sufficiently close to the lane in the pilot test, 0 m, which is the intervention at all times, and 0.4 m, which is the median value. A combination of these two independent variables yielded a total of nine treatment conditions, as shown in Fig.~\ref{fig:2}. A balanced Latin Square was used to avoid the transfer, learning, and fatigue effects.

\begin{figure}
  \begin{center}
  \includegraphics[width=0.5\linewidth]{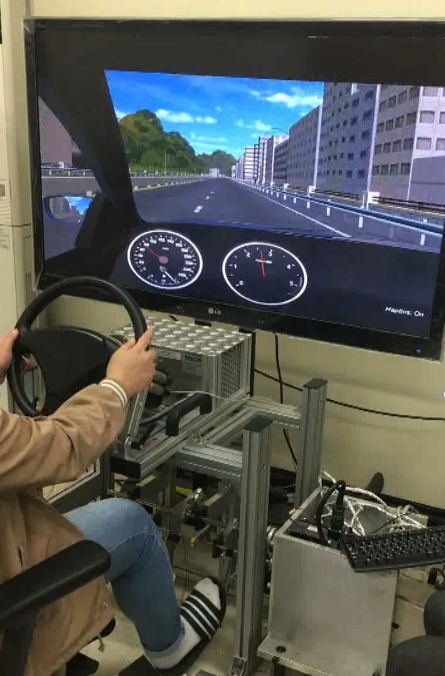}
  \end{center}
  \caption{Driving simulator used in this study.}
  \label{fig:3}
\end{figure}

\subsection{Apparatus and experimental setting}

The simulator consists of a 55-inch display, steering wheel, accelerator pedal, brake pedal, driving seat, and speaker (Fig.~\ref{fig:3}). The distance from the display to the seat is set to approximately 1.2 m for a comfortable field of view of 60$^\circ$. For the LKAS control torque, a commercial steering handle (SENSO-Wheel SD-LC) manufactured by SensoDrive was employed. The virtual environment for the driving task was implemented in Unity 3D with the commercial physics engine of the Vehicle Physics Pro. For realistic simulations, the physical parameters of the actual vehicle such as the mass, dimension, gear ratio, and engine power curves were chosen \citep{lee2018}.

\subsection{Procedure and task}

The participants were first informed about the experiment and provided their consent and demographic information; subsequently, a practice driving session of 5 km was conducted to adapt the participants to the simulator environment. The experiment was conducted on nine counter-balanced treatment conditions; after the driving for each condition, the participants were asked to respond to the questionnaire including the self-reported measures (emotion, trust, disturbance, and overall satisfaction).

The task performed by the participants during the experiment is to drive at a speed of approximately 70 km/h in a 1.5-km straight highway. Each treatment condition comprises two sessions. In the first session, the subject participated in the experiment as an evaluator. They aim to establish their mental model for the control strategy through the torque generated by the LKAS, assuming both normal and careless situations. The steering wheel could be freely manipulated such that the torque could be fully felt. In the second session, the participants participated as an ordinary driver with the distracted situation. They verified the incoming message on the mobile phone and typed the exact information. When the vehicle departs from the lane, the experimenter informs the subject to place the vehicle in the lane. After retaining the vehicle in the lane for at least 3 s, the drivers were asked to continue the typing task.

\subsection{Dependent measures}

As subjective measures for the driving experience, the emotion, trust, disturbance, and overall satisfaction for each treatment condition were collected by self-reported measures on a continuous scale between 0 and 100. The measure for the emotion consisted of valence and arousal in Russell’s Circumplex Model \citep{russell1980circumplex} with the Self-Assessment Manikins \citep{lang1980}, which is a representative assessment for the emotion. As objective measures for the driving performance and activity, three variables were analyzed using the trajectory log data: 

\begin{itemize}
\item Standard deviation of lane position (SDLP): It is estimated as the standard deviation of the lateral position from the lane center. If this value is high, it is interpreted that the lateral driving is not stable.
\item Steering reversal rate (SRR): It is measured by the frequency of steering wheel reversals (corrections) larger than a pre-defined angle \citep{ostlund2005}. In this study, the steering angle change of more than 1$^\circ$ per second was measured.
\item Root mean square of lateral speed (RMSLS): It is estimated as the root mean square of the lateral speed. The lower this value, the better is the driving performance.
\end{itemize}

\section{Results}

All the measures are analyzed using repeated measure ANOVA tests. If the Mauchly sphericity test violates the sphericity assumption, the Greenhouse–Geisser Epsilon \citep{greenhouse1959methods} is used for the analysis. Bonferroni’s post-hoc test is performed if the primary effect or any interaction effect is found to be significant. For all statistical comparisons, $\alpha$ = 0.05 is used as the criterion for statistical significance.

\begin{table*}
\begin{center}
\small\begin{tabular} {llllll}
\hline
Measure & Effect & $df_1$ & $df_2$ & F-value & p-value \\
\hline
Valence         & TOR &  1.295  & 22.023    & 6.112 & p =.016$^*$ \\
                & DEV &  2      & 34        & 4.855 & p =.014$^{*}$ \\
  &  TOR $\times$ DEV &  4      & 68        & 0.586 & p =.674 \\
\hline
Arousal  & TOR &  1.406 & 23.902 & 17.252 & p $<$.001$^{***}$ \\
  &  DEV &  2 & 34 & 0.236 & p =.791 \\
  &  TOR $\times$ DEV &  4 & 68 & 0.685 & p =.605 \\
\hline
Trust  & TOR &  2 & 34 & 36.165 & p $<$.001$^{***}$ \\
  &  DEV &  2 & 34 & 5.370 & p =.009$^{**}$ \\
  &  TOR $\times$ DEV &  4 & 68 & 0.761 & p =.554 \\
\hline
Disturbance  & TOR &  2 & 34 & 32.612 & p $<$.001$^{***}$ \\
  &  DEV &  2 & 34 & 1.521 & p =.233 \\
  &  TOR $\times$ DEV &  4 & 68 & 0.868 & p =.488 \\
\hline
Satisfaction  & TOR &  1.488 & 25.291 & 22.080 & p $<$.001$^{***}$ \\
  &  DEV &  2 & 34 & 6.308 & p =.005$^{**}$ \\
  &  TOR $\times$ DEV &  4 & 68 & 0.801 & p =.529 \\
\hline
SDLP  & TOR &  1.513 & 25.728 & 10.247 & p =.001$^{**}$ \\
  &  DEV &  2 & 34 & 26.896 & p $<$.001$^{***}$ \\
  &  TOR $\times$ DEV &  4 & 68 & 1.428 & p =.234 \\
  \hline
SRR  & TOR &  2 & 34 & 18.724 & p $<$.001$^{***}$ \\
  &  DEV &  2 & 34 & 0.205 & p =.816 \\
  &  TOR $\times$ DEV &  4 & 68 & 2.263 & p =.071 \\
  \hline
RMSLS  & TOR &  2 & 34 & 0.242 & p =.786 \\
  &  DEV &  2 & 34 & 7.395 & p =.002$^{**}$ \\
  &  TOR $\times$ DEV &  2.035 & 34.595 & 2.643 & p =.085 \\
    
\hline
\end{tabular}
\end{center}
\caption{ANOVA tests on all the dependent measures. ($^*$: p $<$ .05, $^{**}$: p $<$ .01, $^{***}$: p $<$ .001) \newline}
\label{tab:overall}
\end{table*}

\begin{figure}
  \begin{center}
  \subfloat[][Effects of TOR]{
    \includegraphics[width=0.45\linewidth]{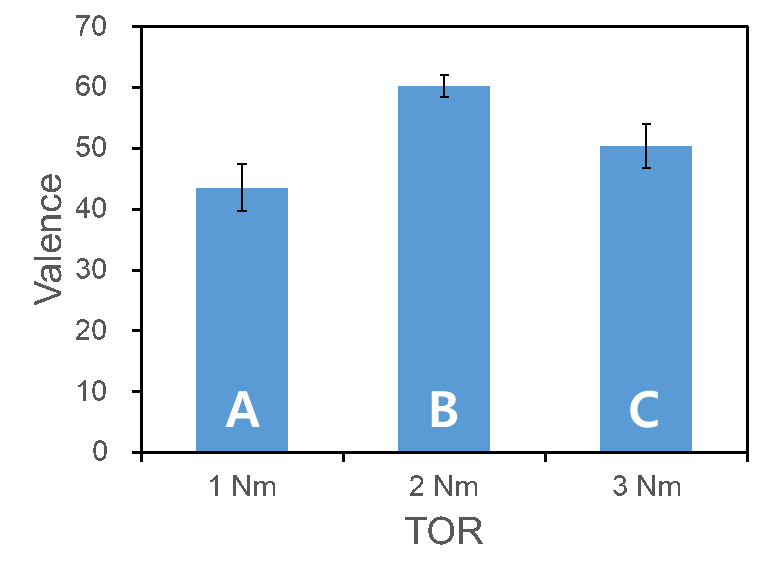}
  }
  \subfloat[][Effects of DEV]{
    \includegraphics[width=0.45\linewidth]{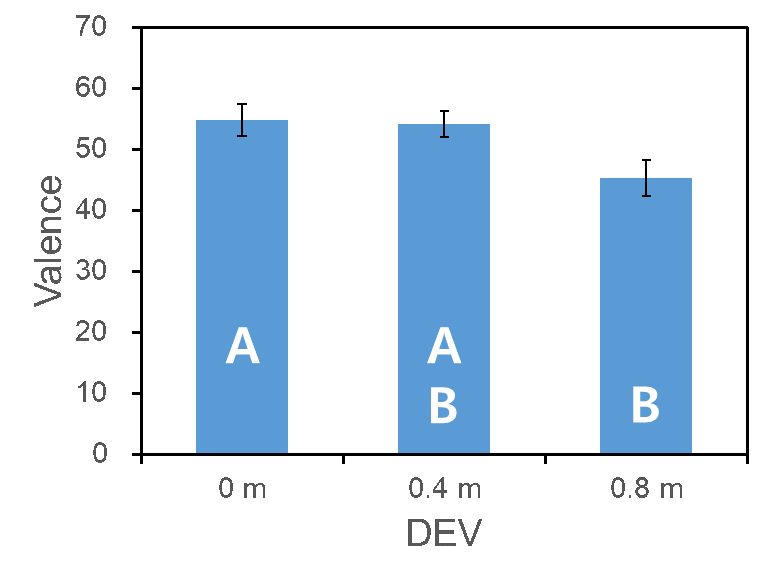}
  }
  \end{center}
  \caption{Primary effects of TOR and DEV on valence of emotion. Error bar = SEM. The same alphabet characters indicate no significant difference by Bonferroni post-hoc comparison with $\alpha$ =.05.}
  \label{fig:4}
\end{figure}

\begin{figure}
  \begin{center}
    \includegraphics[width=0.5\linewidth]{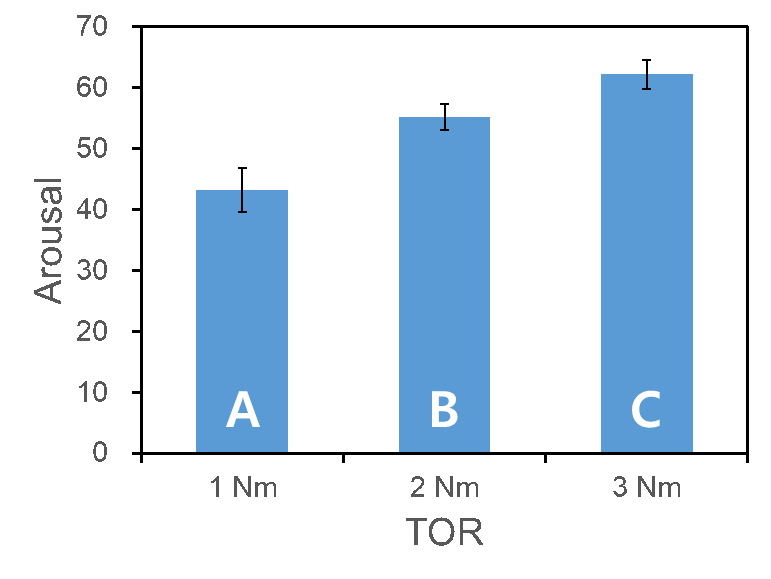}
  \end{center}
  \caption{Primary effects of TOR on arousal of emotion. Error bar = SEM. The same alphabet characters indicate no significant difference by Bonferroni post-hoc comparison with $\alpha$ =.05.}
  \label{fig:5}
\end{figure}

\subsection{Subjective measures (Driving experience)}

\subsubsection{Emotion}

The primary effects of TOR and DEV are found on the valence of emotion (Table \ref{tab:overall}). Post-hoc analysis shows that the valence is the highest in the 2-Nm TOR condition (60.3 points), followed by 3-Nm TOR (50.5 points), and is the highest in 0-m DEV (54.8 points). However, no significant difference is observed in the 0-m DEV and 0.4-m DEV, as well as in the 0.4-m DEV and 0.8-m DEV (Fig.~\ref{fig:4}). No interaction effect occurred between TOR and DEV (p =.674).

The arousal of emotion shows a significant difference across TORs (Table \ref{tab:overall}). The arousal of drivers does not differ significantly across DEVs. The post-hoc analysis shows that the arousal is the highest in the 3-Nm TOR condition (62.2 points), followed by 2-Nm TOR (55.2 points) (Fig.~\ref{fig:5}). No interaction effect occurs between TOR and DEV (p =.605). In summary, a large TOR increases the arousal of emotion. 

\begin{figure}
  \begin{center}
    \subfloat[][Effects of TOR]{
    \includegraphics[width=0.45\linewidth]{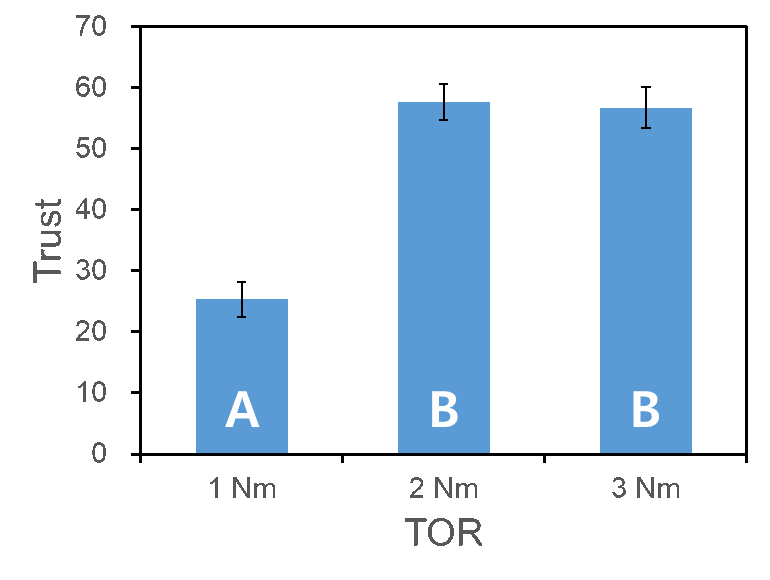}
  }
  \subfloat[][Effects of DEV]{
    \includegraphics[width=0.45\linewidth]{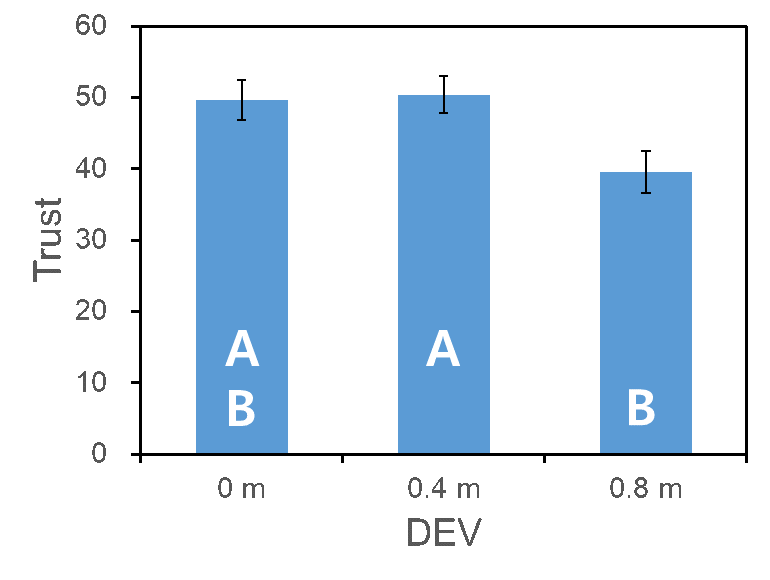}
  }
  \end{center}
  \caption{Primary effects of TOR and DEV on the trust. Error bar = SEM. Error bar = SEM. The same alphabet characters indicate no significant difference by Bonferroni post-hoc comparison with $\alpha$ =.05.}
  \label{fig:6}
\end{figure}

\subsubsection{Trust}

The primary effects of TOR and DEV are found on the trust (Table 1). The trust is the lowest at 1-Nm TOR (25.3 points). The difference of the trust at 2-Nm TOR and 3-Nm TOR is not statistically significant. For the DEV, 0.8 m (39.5 points) was the lowest, followed by 0.0 m and 0.4 m with 49.7 points and 50.4 points, respectively. From the post-hoc analysis, the DEV of 0.0 m and 0.4 m (group A in Fig.~\ref{fig:6}(b)), and 0.0 m and 0.8 m (group B in Fig.~\ref{fig:6}(b)) showed no statistically significant differences. No interaction effect occurs between TOR and DEV (p =.554).

\begin{figure}
  \begin{center}
    \includegraphics[width=0.5\linewidth]{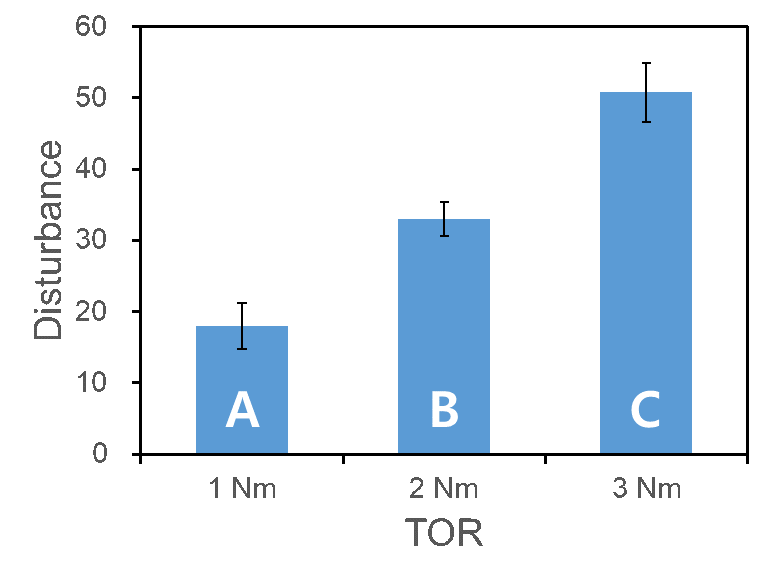}
  \end{center}
  \caption{Primary effects of TOR on the feeling of disturbance. Error bar = SEM. The same alphabet characters indicate no significant difference by Bonferroni post-hoc comparison with $\alpha$ =.05.}
  \label{fig:7}
\end{figure}

\subsubsection{Disturbance}
The feeling of disturbance shows a significant difference across TORs (Table \ref{tab:overall}). The disturbance does not differ significantly across DEVs. The post-hoc analysis shows that the disturbance is the highest in the 3-Nm TOR condition (50.8 points), followed by 2-Nm TOR (33.0 points), and 1-Nm TOR (18.0 points), as shown in Fig.~\ref{fig:7}. No interaction effect occurs between TOR and DEV (p =.488). In summary, a large TOR increases the feeling of disturbance.

\subsubsection{Satisfaction}
\begin{figure}
  \begin{center}
    \subfloat[][Effects of TOR]{
    \includegraphics[width=0.45\linewidth]{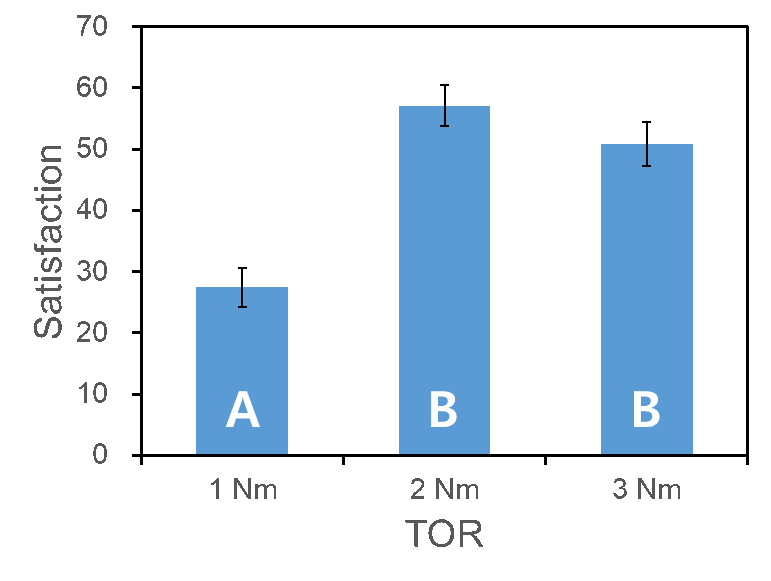}
  }
  \subfloat[][Effects of DEV]{
    \includegraphics[width=0.45\linewidth]{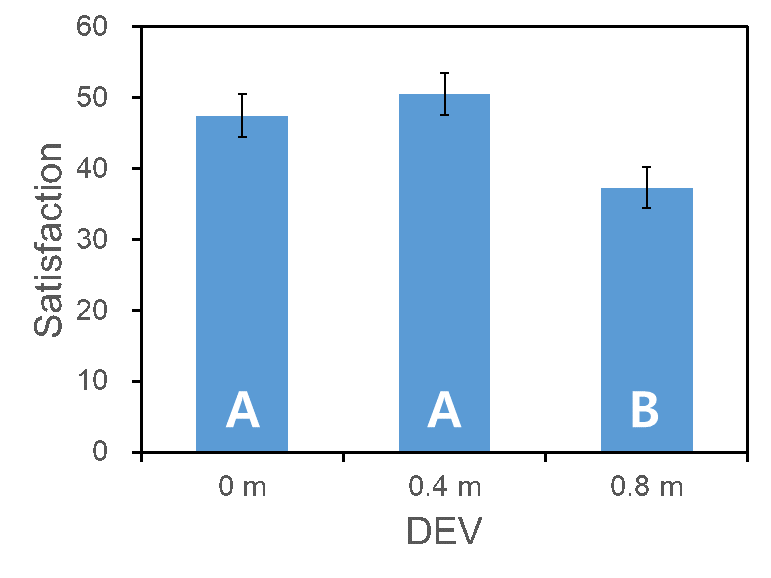}
  }
  \end{center}
  \caption{Primary effects of TOR and DEV on the satisfaction. Error bar = SEM. Error bar = SEM. The same alphabet characters indicate no significant difference by Bonferroni post-hoc comparison with $\alpha$ =.05.}
  \label{fig:8}
\end{figure}

The TOR and DEV significantly affect the overall satisfaction (Table \ref{tab:overall}). No interaction effect occurs between TOR and DEV (p =.529). As shown in Fig.~\ref{fig:8}, the satisfaction is the highest at 2-Nm TOR with 57.1 points, followed by 3-Nm TOR (50.8 points), and 1-Nm TOR (27.4 points). No significant difference is observed in 2-Nm TOR and 3-Nm TOR. In terms of DEV, 0.4 m records the highest satisfaction, (50.6 points). The 0.0 m DEV and 0.8 m DEV record 47.5 points and 37.3 points, respectively. No statistical difference occurs between 0.0-m DEV and 0.4-m DEV in the post-hoc analysis. 

\begin{figure}
  \begin{center}
    \subfloat[][Effects of TOR]{
    \includegraphics[width=0.45\linewidth]{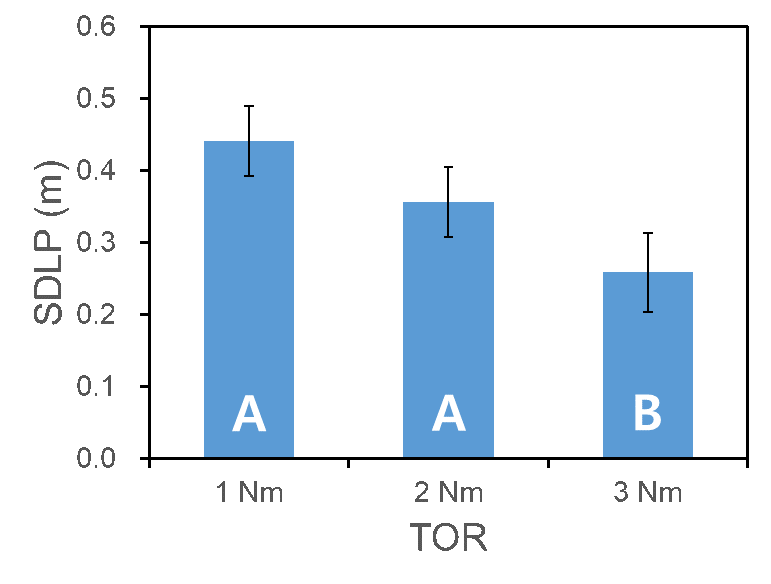}
  }
  \subfloat[][Effects of DEV]{
    \includegraphics[width=0.45\linewidth]{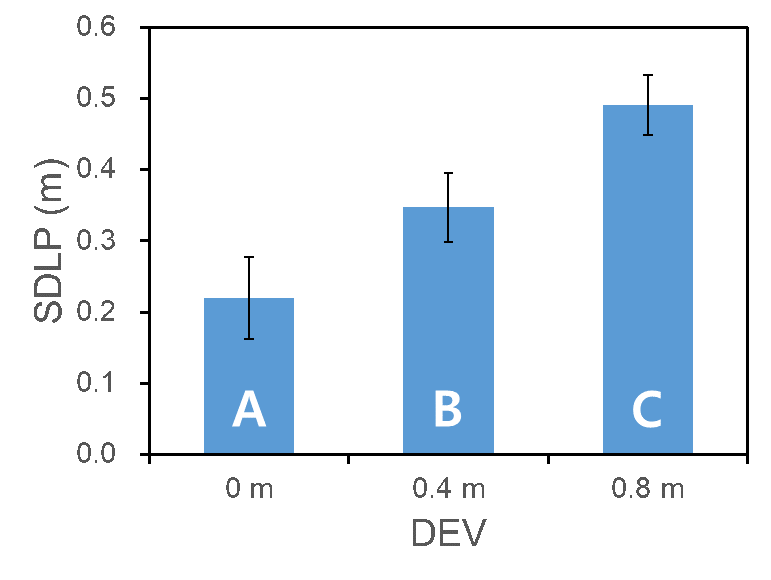}
  }
  \end{center}
  \caption{Primary effects of TOR and DEV on standard deviation of lane position (SDLP). Error bar = SEM. Error bar = SEM. The same alphabet characters indicate no significant difference by Bonferroni post-hoc comparison with $\alpha$ =.05.}
  \label{fig:9}
\end{figure}

\begin{figure}
  \begin{center}
    \includegraphics[width=0.5\linewidth]{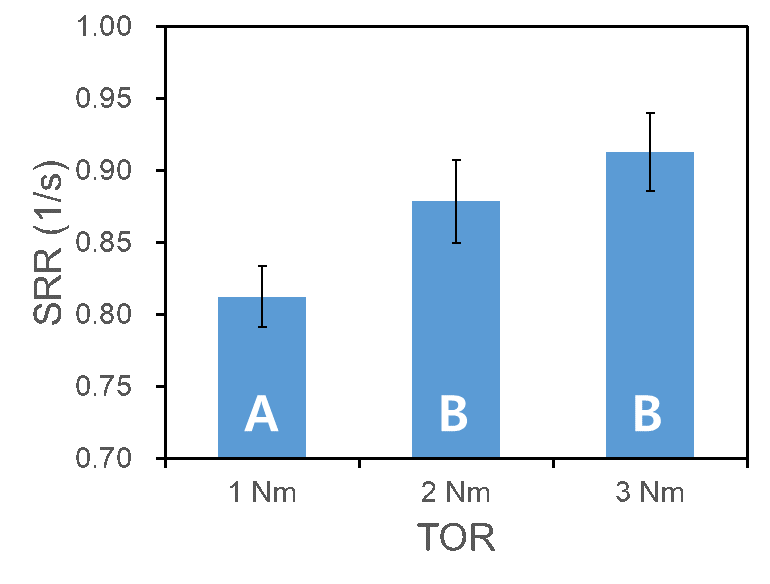}
  \end{center}
  \caption{Primary effects of TOR on steering reversal rate (SRR). Error bar = SEM. The same alphabet characters indicate no significant difference by Bonferroni post-hoc comparison with $\alpha$ =.05.}
  \label{fig:10}
\end{figure}

\begin{figure}
  \begin{center}
    \includegraphics[width=0.5\linewidth]{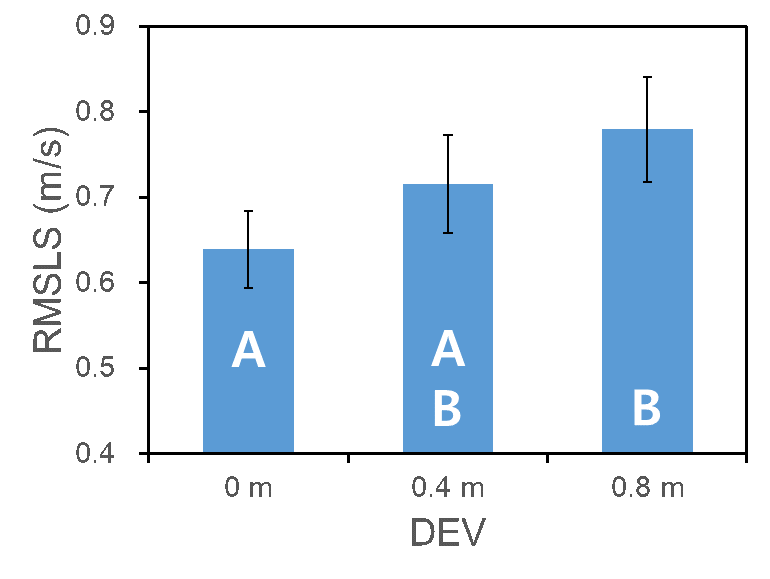}
  \end{center}
  \caption{Primary effects of TOR on root mean square of lateral speed (RMSLS). Error bar = SEM. The same alphabet characters indicate no significant difference by Bonferroni post-hoc comparison with $\alpha$ =.05.}
  \label{fig:11}
\end{figure}

\subsection{Objective measures (Driving behavior)}

\subsubsection{Standard deviation of lane position (SDLP)}
As shown in Table \ref{tab:overall}, the primary effects of TOR and DEV are found on the SDLP. The post-hoc analysis shows that the SDLP is the lowest in the 3-Nm TOR condition (0.26 m), followed by 2-Nm TOR (0.36 m), and 1-Nm TOR (0.44 m), and is the lowest in 0.0-m DEV (0.22 m), followed by 0.4-m DEV (0.35 m), and 0.8-m DEV (0.49 m). No significant difference is observed in 1-Nm TOR and 2-Nm TOR (Fig.~\ref{fig:9}). No interaction effect occurs between TOR and DEV (p =.234).

\subsubsection{Steering reversal rate (SRR)}
The SRR shows a significant difference across the TORs (Table \ref{tab:overall}). The SRR does not differ significantly across DEVs. The post-hoc analysis shows that the SRR is the highest in the 3-Nm TOR condition (0.913 s$^{-1}$), followed by 2-Nm TOR (0.879 s$^{-1}$), and 1-Nm TOR (0.812 s$^{-1}$). No significant difference is observed in 2-Nm TOR and 3-Nm TOR (Fig.~\ref{fig:10}). No interaction effect occurs between TOR and DEV (p =.071).
 
\subsubsection{Root mean square of lateral speed (RMSLS)}
The RMSLS shows a significant difference across the DEVs (Table \ref{tab:overall}). It does not differ significantly across the TORs. The post-hoc analysis shows that the RMSLS is the highest in the 0.8-m DEV condition (0.78 m/s), followed by 0.4-m DEV (0.72 m/s), and 0.0-m DEV (0.64 m/s). From the post-hoc analysis, the DEV of 0.0 m and 0.4 m (group A in Fig.~\ref{fig:11}), and 0.4 m and 0.8 m (group B in Fig.~\ref{fig:11}) showed no statistically significant differences. No interaction effect occurs between TOR and DEV (p =.085).

\section{Discussion}

\subsection{Primary effect}

The experiment provided several meaningful points in terms of the primary effect, TOR, and DEV. The feeling of disturbance tends to increase by increasing the TOR. This implies that the torque of the LKAS can cause the driver to disturb or interfere. However, weak interventions that do not cause disturbance are unacceptable. The high SDLP at 1-Nm TOR implies the poor performance of the control strategy. The SRR is the lowest at 1-Nm TOR. In general, a low SRR can be interpreted as a small workload; however, in this study, it should be understood as a correction of the LKAS in the same level of visual distraction. Therefore, it can be interpreted that the frequency of correction of the LKAS is low at 1-Nm TOR, which eventually led to a high SDLP. Because the drivers were aware of this performance defect, the condition with 1-Nm TOR appears to reduce the trust and overall satisfaction. Meanwhile, 3-Nm TOR showed the lowest SDLP and high level of trust and satisfaction score. However, the valence of emotion was negative and the disturbance was high. In the 3-Nm TOR, many participants responded that they felt that the LKAS was overly interfering with their normal driving. These factors appear to affect overall satisfaction and yielded a slightly lower satisfaction score than 2-Nm TOR. In the control strategy with 2-Nm TOR, the participants actually showed the most positive emotions and the high score of overall satisfaction. Thus, the 2-Nm TOR is recommended within the experimental conditions. Among the dependent variables, the SRR showed a tendency to increase as the TOR increased. 

Within the experimental conditions in terms of the DEV, the gap to prevent excessive interventions does not significantly affect the driving experience at 0.4 m; instead, it caused the driving experience to deteriorate at 0.8 m. Most participants responded that the intervention of torque started too late at 0.8 m and that the system did not perform properly, which eventually led to the low trust and satisfaction scores. In the 0.8-m condition, many cases were observed in which the drivers perceived a lane departure as late, or recognized the correction torque and then suddenly operate the steering wheel. These observations support unstable driving with a high lateral speed at 0.8 m. No difference was observed in the driving experience between 0.0 m and 0.4 m, but a difference in the driving performance was observed. The lowest SDLP at 0.0 m shows that a stable lane keeping performance is possible through sustained and continuous torque intervention. Thus, active lane keeping (0.0-m DEV) is recommended within the experimental conditions.

\begin{figure}
  \begin{center}
    \includegraphics[width=0.7\linewidth]{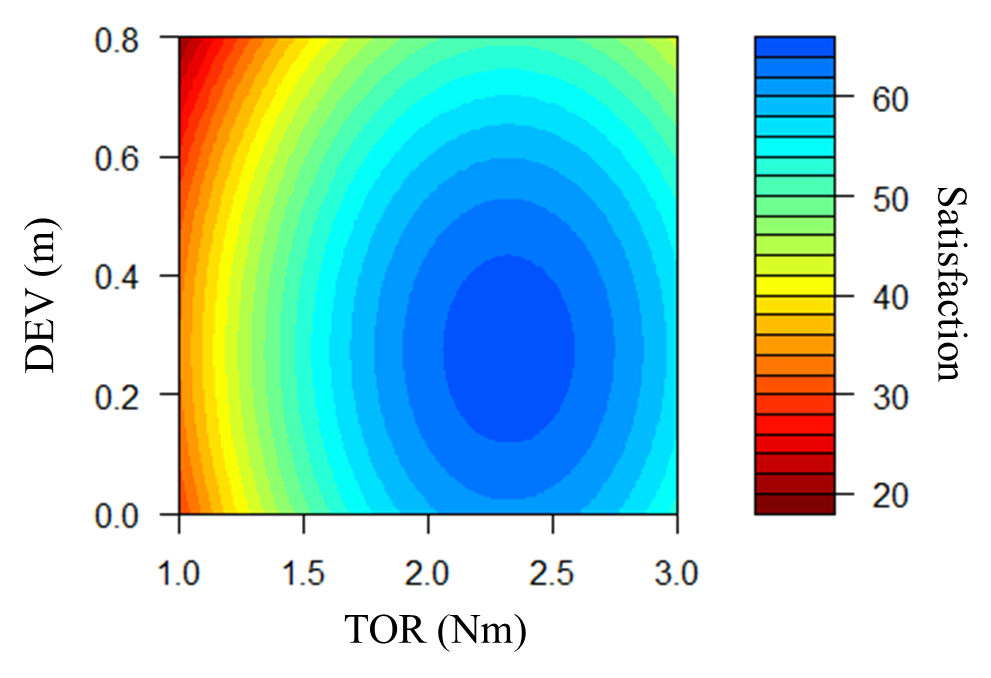}
  \end{center}
  \caption{Contour plot for the selected model of overall satisfaction.}
  \label{fig:12}
\end{figure}

\subsection{Regession analysis}

To obtain the combination of optimal TOR and DEV values for outstanding satisfaction, the best subsets approach, which aims to obtain the best-fit regression model from all possible subset models \citep{hosmer2013applied}, was used with the polynomial model given in eq.~\ref{eq:3} below:
\begin{equation}
    \label{eq:3}
    y = \sum_{i=1}^na_ix_i + \sum_{i=1}^n\sum_{j=1}^nb_{ij}x_ix_j + \sum_{i=1}^nc_ix^2_i
\end{equation}
where $y$ is a dependent variable; $x_i$ is the $i$th independent variable; $a$, $b$, and $c$ are the coefficients of each corresponding variable. Consequently, the model for satisfaction is derived as eq.~\ref{eq:4}.
\begin{equation}
    \label{eq:4}
    \begin{split}
    SAT = -18.01 TOR^2-50.93 DEV^2+83.75 TOR \\
    +28.01 DEV-35.96
    \end{split}
\end{equation}
The Mallows C-p value was 4.9 and the adjusted R$^2$ was 0.942. We confirmed that the appropriate TOR and DEV are required for the best satisfaction, according to the contour plot of the model shown in Fig.~\ref{fig:12}. The TOR of approximately 2.33 Nm and DEV of approximately 0.27 m are the most satisfactory combinations of the parameters.

\subsection{Limitations}

The present study is one of the first attempts to thoroughly examine the effects of the LKAS design parameters on the driving experience including emotion. However, this study has the following limitations. As a limitation of the driving simulator used in this study, the perception of the rotational inertia owing to lateral control was not considered. Only the self-reported measure was used to collect the driving experience. It will be able to increase the validity of the research by adding other objective measures, such as the use of electroencephalography (EEG) in relation to valence and arousal of emotions \citep{schmidt2001frontal, bos2006eeg, liu2010real}. 

\section{Conclusion and Future works}

In this study, we investigated how driving experience changed according to various steering control strategies of the LKAS, and suggested a strategy with the optimal parameter. Hence, a total of nine LKAS control strategies were created in the driving simulator, combining three TOR and three DEV conditions. Eighteen participants evaluated each strategy regarding emotion, trust, disturbance, and overall satisfaction, and the trajectory log data of distracted driving was analyzed in terms of driving performance. The result of this experiment showed that 1-Nm TOR caused the degradation in lane keeping performance, and therefore did not provide the driver with sufficient trust and satisfaction. The TOR of 3 Nm showed excellent lane keeping performance, but the feeling of disturbance and valence of emotion were not positive. The DEV of 0.8 m condition was recognized as too late to properly control the steering wheel. Regarding the 1-Nm TOR, it did not provide the driver with sufficient trust and satisfaction. The driving experience of 0.0-m DEV and 0.4-m DEV was similar, but the 0.0-m condition was recommended because of the better driving performance. For the best satisfaction, the optimal parameters derived from the regression model were approximately 2.32-Nm TOR and approximately 0.27-m DEV.

The following research will be conducted in the future. To verify the effects of gender, we will conduct experiments with the same number of male and female participants. From our observations, we suspect that the driver is more likely to focus on the secondary task for a trustworthy LKAS strategy. Thus, the analysis of the secondary task should be performed. We will also employ EEG to measure the driver’s distraction. The validation of other important LKAS parameters should be performed. One example is torque reduction, which reduces the intervention torque of the LKAS at a certain rate when the driver applies more than a certain amount of torque to the steering wheel. We will verify the effect of torque reduction on the feeling of disturbance and overall satisfaction.

\section*{References}

\bibliography{mybibfile}

\end{document}